\documentstyle[aps,12pt]{revtex}

\newcommand{\ba}{\begin{eqnarray}}
\newcommand{\ea}{\end{eqnarray}}
\newcommand{\baa}{\begin{array}}
\newcommand{\eaa}{\end{array}}

\newcommand{\beq}{\begin{equation}}
\newcommand{\eeq}{\end{equation}}

\newcommand{\Cg}{{\mathcal C}_g}

\newcommand{\tr}{{\mathrm{tr}}}
\newcommand{\hps}{h_{\phi}}
\newcommand{\hsc}{h_{\sigma}}
\newcommand{\sigmaI}{\sigma_{\mathrm{I}}}
\newcommand{\sigmaII}{\sigma_{\mathrm{II}}}
\newcommand{\sigmaIII}{\sigma_{\mathrm{III}}}

\newcommand{\comment}[1]{}

\makeatletter
\renewcommand{\@makefnmark}{\hbox{\mathsurround=0pt} $^{\@thefnmark)}$}
\renewcommand{\@makefntext}[1]{\parindent=1em\noindent
\hbox to 1.8em{\hss $^{\@thefnmark )}$}#1}
\renewcommand{\@biblabel}[1]{#1.\ }
\makeatother

\begin{document}
\large

\title{Strong Decays of Scalar Glueball in a
Scale-Invariant Chiral Quark Model}
\author{M. K. Volkov,
V. L. Yudichev}
\address{
Joint Institute for Nuclear Research,\\
 Dubna, Russia}
\maketitle
\bigskip

\begin{abstract}
An effective meson Lagrangian including a scalar glueball is
constructed on the base of $U(3)\times U(3)$ chiral  symmetry.
The glueball is introduced into the meson Lagrangian by using the
principle of scale invariance of an effective Lagrangian and
the dilaton model. The singlet-octet mixing of scalar meson
states is described by means of 't Hooft interaction. The
contribution of the scalar and pseudoscalar anomalies into the
breaking of scale invariance is taken into account. The mixing
of quarkonia with the glueball is described. The mass spectrum
of scalar mesons together with the glueball
and also their strong decay widths
are calculated. From comparing the obtained results with
experimental data, it follows that $f_0(1500)$ is rather a
glueball, whereas $f_0(1710)$ is a quarkonium. This accords
with the results obtained in our previous works where
radially-excited scalar meson states were described.
It is shown that $\rho$-mesons play an important role in
the description of glueball decays.
\end{abstract}


\newpage

\section{ Introduction}

The self-interaction of gluons, a peculiarity of QCD, gave an idea
that gluons can form bound states that can propagate as particles
in the space. Unfortunately, because of theoretical problems,
there is no yet the exact answer to whether  these states really exist
or not. However, from recent lattice simulations
\cite{Sexton,LeeWeingarten,VaccarinoWeingarten}
one can conclude
that it is most probably that glueballs are
real objects of our world.
Having assumed that glueballs exist,
one can try to construct a model to describe their interaction with other
mesons, their properties, such as,
e.~g., mass and decay width, and to identify them with observed
resonances.

An exact microscopic description of bound gluon states cannot be done
systematically in the framework of QCD. In this situation,
QCD-motivated phenomenological models are the tool that can
help to deal with glueballs as well as with quarkonia which form the most
of observed meson states. However, using these models to describe glueballs,
we encounter many difficulties concerning, e.~g., the ambiguity of the
ways of including glueballs into models and identification of
experimentally observed meson states.
This explains the variety of points of view on this problem.

First of all, we do not know the exact mass of a glueball.
From the quenched QCD lattice simulations, Weingarten
(see, e.~g., \cite{Sexton,VaccarinoWeingarten}) concluded that
the lightest scalar glueball is expected around 1.7 GeV. Amsler \cite{Amsler}
considered
the state $f_0(1500)$ as a candidate for the scalar glueball.
QCD sum rules \cite{Naris_98} and $K$-matrix method \cite{Aniso_98}
showed that both $f_0(1500)$ and $f_0(1710)$ are mixed states with
large admixture of the glueball component.

All  bound isoscalar $q\bar q$ states are subject to mixing with
glueballs, and their  spectrum  has many interpretations made
by different authors.
For instance, Palano \cite{Palano}  suggested a scenario, in
which the states $a_0(980)$, $K_0^*(1430)$, $f_0(980)$, and $f_0(1400)$
form a nonet. The state $f_0(1500)$ is considered as the scalar glueball.
T\"ornqvist \textit{et al.\/}~\cite{Tornqvist} looked upon the states $f_0(980)$ and $f_0(1370)$
as manifestations of the ground and excited $s\bar s$ states, and upon
the state $f_0(400-1200)$
as the ground $u\bar u$ state.
 Eef van Beveren
\textit{et al.\/}~\cite{Beveren} considered the states  $f_0(400-1200)$ and $f_0(1370)$ as
ground $u\bar u$ states,
and the states $f_0(980)$ and $f_0(1500)$  as ground $s\bar s$ states.
Two states for each $q\bar q$ system occur due to pole doubling, which
takes place for scalar mesons in their model.
Shakin \textit{et al.\/}~\cite{Shakin,Shakin:2000} obtained from a nonlocal confinement model that
the $f_0(980)$ resonance is the ground $u\bar u$ state, and $f_0(1370)$
is the ground $s\bar s$ state. The state $f_0(1500)$ is considered as a radial
excitation of $f_0(980)$. They believe the mass of  scalar glueball to be
1770 MeV.

In our recent papers \cite{ECHAJA1999},
following  methods given in \cite{Volk_86,Volk_82,Weiss,YaF},
we showed that all experimentally observed scalar meson states with
masses in the interval from 0.4 to 1.71 GeV can be interpreted
as members of two scalar meson nonets ---
the  ground state of the meson nonet (lighter than 1 GeV)  and its first
radial excitation (heavier than 1 GeV).
We considered all scalar mesons as $q{\bar q}$ bound
states and took into account the singlet-octet mixing  caused by
't Hooft interaction. In \cite{ECHAJA1999}, we obtained
a scalar isoscalar state with mass 1600 MeV and had to choose
to which of the experimentally observed
states, $f_0(1500)$ or $f_0(1710)$, we should ascribe it.
From our analysis of the strong decay rates calculated in our
model we found that $f_0(1710)$ fits to the nonet of quarkonia better than $f_0(1500)$.
Therefore, we supposed that the state $f_0(1500)$ contains
greater admixture of the scalar
glueball (see \cite{Naris_98,Aniso_98}). However,
the final decision should be made after including the scalar
glueball into the model and taking  account of its mixing
with quarkonia.
In the present work, that is devoted to solving this problem,
from the analysis of strong decay widths of the glueball we again
come to an analogous conclusion\footnote{
However, radially-excited states are not yet considered.}.

To describe the properties of the glueball and its interaction with
quarkonia, one should introduce
a scalar isoscalar dilaton field $\chi$ into our model,
in addition to the quarkonia that
have  already been described~\cite{ECHAJA1999}.
For this purpose, one can make use of the idea of approximate
scale invariance of effective Lagrangians based on the dilaton
model.
Such models were studied by many authors
(see, e.~g., \cite{gb:models6,Jami_96,Kusa_93,Andr_86,Elli_84}).
Unfortunately, there is no unique way to introduce
the dilaton field into a chiral  Lagrangian.
This explains the large number of models dealing with
glueballs.

The guideline one should follow when introducing
the dilaton field into an effective meson Lagrangian is
to reproduce  the Ward identity connected
with the scale anomaly. The latter leads to the following equation
for the vacuum expectation value of the divergence of the
dilatation current:
 \beq
 \label{Ward}
 \langle\partial_{\mu}S^{\mu}\rangle=\Cg-\sum_{q=u,d,s}m_{q}^0
 \langle\bar qq\rangle,
 \eeq
 \beq
 \Cg=\left({11\over 24}N_c - {1\over 12}N_f \right)
 \left\langle{\alpha\over \pi}G^2_{\mu\nu}\right\rangle, \label{gluoncon}
 \eeq
where $N_c$ is the number of colours; $N_f$ is
the number of flavours;  $\langle {\alpha\over\pi}G_{\mu\nu}^2\rangle$ and
$\langle\bar qq\rangle$ are
the gluon and quark condensates; $m^0_{q}$ is the current quark mass.

In this paper we are going to use the most natural method of
introducing the dilaton field into the effective Lagrangian
by requiring that, in the chiral limit, our Lagrangian should
be scale-invariant except for the dilaton potential
and terms induced by gluon anomalies.  To
realize this program, one should multiply all dimensional
parameters of the original Lagrangian (without dilaton) by
a corresponding power of
the dilaton field divided by its vacuum expectation value
$\chi_c$.
Thus, instead of the four-quark
coupling constant $G$, the 't Hooft coupling constant $K$,
ultraviolet cutoff $\Lambda$ (necessary for
regularizing the divergent integrals coming from quark
loops), and the constituent quark masses $m_q$ $(q=u,s)$, one
should use $G(\chi_c/\chi)^2$, $K(\chi_c/\chi)^5$,
$\Lambda(\chi/\chi_c)$, and $m_q(\chi/\chi_c)$.

Current quark masses $m^0_q$ are not multiplied by the
dilaton field and violate scale invariance explicitly, as it
takes place in QCD.  Their contribution to the divergence of
dilatation current is determined by quark condensates
and disappears in the chiral limit (see (\ref{Ward})).

The scale invariance is also broken by those terms in the
effective Lagrangian that are induced by the pseudoscalar and
scalar gluon anomalies and look as follows \cite{schechter,volk82}
\beq
L_{\mathrm an}=-\hps\phi_0^2+\hsc\sigma_0^2, \label{anomalynotscaled}
\eeq
where
 $\hps, \hsc $ are constants.
 $\phi_0=\sqrt{2/3}\phi_u-\sqrt{1/3}\phi_s$,
$\phi_0$ and $\sigma_0\quad (\langle\sigma_0\rangle \not =0)$
$\sigma_0=\sqrt{2/3}\sigma_u-\sqrt{1/3}\sigma_s$, where
$\sigma_u$ ($\langle\sigma_u\rangle\not=0$) consists of $u(d)$-quarks
and $\sigma_s$ ($\langle\sigma_s\rangle\not=0$) of $s$-quarks.

These terms appear due to the 't Hooft  interaction.
When restoring scale invariance of the effective Lagrangian by
inserting dilaton fields (the procedure of  the restoration
of scale invariance is given in Sect.~3),
these terms must be treated separately.
Moreover, it turns out that these terms determine the most of
quarkonia-glueball mixing.

Omitting, for a moment, the 't Hooft interaction
in our approach, we require
the Lagrangian to be scale-invariant in the chiral limit  both
before and after the spontaneous breaking of chiral symmetry (SBCS),
except for  the dilaton potential.
This property can be obtained by considering
(after bosonization when the effective Lagrangian is expressed
in terms of bosonic scalar and pseudoscalar
fields $\sigma$ and $\phi$) the shift of the scalar meson field
$\sigma$
\beq
\sigma=\sigma'-m\frac{\chi}{\chi_c}, \qquad (m^0=0), \label{sigma:shift}
\eeq
where $\langle\sigma'\rangle_0=0,\quad \langle\sigma\rangle_0=-m$,
guaranteeing that the relation (\ref{Ward}) is satisfied.
The nonzero vacuum expectation value of $\sigma$ appears as a result
of SBCS, and thus, the constituent quark mass $m$ is produced.
In the case of nonvanishing current quark masses,
 (\ref{sigma:shift}) changes by including an additional
(nonscaled) mass term
$m^0$ into the r.h.s.
\beq
\sigma=\sigma'-m\frac{\chi}{\chi_c}+m^0. \label{sigma:shift:2}
\eeq

The structure of the paper is as follows. In Section 2, we derive the
usual $U(3)\times U(3)$-flavour symmetric effective Lagrangian with
the 't Hooft interaction and without dilaton fields.  In Section 3,
the dilaton field is introduced into the effective Lagrangian obtained
in Section 2.  In Section 4, the gap equations are investigated,
the quadratic (in fields) terms are deduced and the
mixing matrix for scalar isoscalar states is introduced.  In
Section 5,  the numerical estimates for  the model parameters
are given. The main strong
decays of scalar isoscalar mesons are calculated in Section 6.
It is shown there that $\rho$-mesons play an important role in
the decay of a glueball into four pions.
Finally,
in the Conclusion, we discuss the obtained results.

\section{Chiral effective Lagrangian with  't Hooft interaction}

A $U(3)\times U(3)$ chiral Lagrangian with the 't Hooft interaction
was investigated in paper \cite{Cimen_99}.  It consists of three terms
(see below).  The first term represents the free
quark Lagrangian, the second is composed of four-quark vertices as in
the NJL model, and the last one describes the six-quark 't Hooft
interaction~\cite{Dorokhov92} that is necessary to solve the $U_A(1)$
problem.
\ba
L& =& {\bar q}(i{\hat \partial} - m^0)q + {G\over 2}\sum_{a=0}^8
[({\bar q} {\lambda}_a q)^2 +({\bar q}i{\gamma}_5{\lambda}_a q)^2]
-\nonumber\\ &&- K \left\{ {\det}[{\bar q}(1+\gamma_5)q]+{\det}[{\bar
q}(1-\gamma_5)q]
\right\}.
\label{Ldet}
\ea
Here $G$ and $K$ are coupling constants, $\lambda_a\; (a=1,...,8)$ are
the Gell-Mann matrices, $\lambda_0 = {\sqrt{2/ 3}}$~{\bf 1}, with {\bf
1} being the unit matrix; $m^0$ is a current quark mass matrix with
diagonal elements $m^0_u$, $m^0_d$, $m^0_s$ $(m^0_u \approx m^0_d)$.

The standard bosonization procedure
for local quark models consists in replacing
the four-quark vertices by Yukawa coupling of quarks with bosonic
fields which enables one to perform the integration over quark fields.
The final effective bosonic Lagrangian appears
then as a result of the calculation of the quark determinant.
To realize this program, it is necessary,
using the method described in
\cite{Cimen_99,Dorokhov92,Vogl_91,Kleva_92}, to go from
Lagrangian (\ref{Ldet})
to an intermediate Lagrangian which contains only four-quark vertices
\ba
&&L = {\bar q}(i{\hat \partial} -
\overline{m }^0)q + {1\over 2}\sum_{a,b=1}^9[G_{ab}^{(-)} ({\bar q}{\tau}_a
q)({\bar q}{\tau}_b q) + G_{ab}^{(+)}({\bar q}i{\gamma}_5{\tau}_a
q)({\bar q}i{\gamma}_5{\tau}_b q)], \label{LGus}
\ea
where
\ba &&{\tau}_a={\lambda}_a
~~~ (a=1,...,7),~~~\tau_8 = ({\sqrt 2} \lambda_0 + \lambda_8)/{\sqrt
3},\nonumber\\ &&\tau_9 = (-\lambda_0 + {\sqrt 2}\lambda_8)/{\sqrt 3},
\nonumber \\
&&G_{11}^{(\pm)}=G_{22}^{(\pm)}=G_{33}^{(\pm)}= G \pm
4Km_sI^\Lambda_1(m_s), \nonumber \\
&&G_{44}^{(\pm)}=G_{55}^{(\pm)}=G_{66}^{(\pm)}=G_{77}^{(\pm)}= G \pm
4Km_uI^\Lambda_1(m_u), \nonumber \\ &&G_{88}^{(\pm)}= G \mp
4Km_sI^\Lambda_1(m_s),\quad G_{99}^{(\pm)}= G,\nonumber\\
&&G_{89}^{(\pm)}=G_{98}^{(\pm)}= \pm 4{\sqrt
2}Km_uI^\Lambda_1(m_u),\nonumber\\&&G_{ab}^{(\pm)}=0\quad (a\not=b; \quad
a,b=1,\dots,7),\nonumber\\
&&G_{a8}^{(\pm)}=G_{a9}^{(\pm)}=G_{8a}^{(\pm)}=G_{9a}^{(\pm)}=0\quad (a=1,\dots, 7),
\label{DefG}
\ea
and $\bar m^0$ is a diagonal matrix composed of modified current quark masses:
\ba
    \overline{m}^0_u&=&m^0_u- 32 K m_u m_s
    I^{\Lambda}_1(m_u)I^{\Lambda}_1(m_s) \label{twoloopcorrect1},\\
    \overline{m}^0_s&=&m^0_s- 32 K m_u^2
    I^{\Lambda}_1(m_u)^2\label{twoloopcorrect2}.
\ea
Here $m_u$ and $m_s$ are constituent quark masses and the integrals
\ba
I^{\Lambda}_n(m_a)={N_c\over (2\pi)^4}\int d^4_e k {\theta (\Lambda^2
-k^2)
\over (k^2 + m^2_a)^n}, \qquad (n=1,2;\; a=u,s),
\label{DefI}
\ea
are calculated in the Euclidean metric and regularized by a simple
$O(4)$-symmetric ultraviolet cutoff $\Lambda$. For
$I^{\Lambda}_1(m_a)$ one gets
\beq
  I^{\Lambda}_1(m_a)=\frac{N_c}{16\pi^2}
\left(\Lambda^2-m_a^2\ln\left({\Lambda^2\over m_a^2}+1\right)\right),
\label{I1}
\eeq
where $m_a$ represents a corresponding constituent quark mass%
: $m_u$ or $m_s$.
Note that we have introduced the notation of
constituent quark mass  already here,
although they will be consistently considered only later,
when discussing mass gap equations (compare (\ref{gap:eq:u})
and (\ref{gap:eq:s}) below) and the
related shift of scalar meson
fields. However, as we want to use an effective four-fermion
interaction instead of the original six-quark one, we have to
use full quark propagators with constituent quark
masses to
calculate quark loop corrections for the constant $G$ (see
(\ref{DefG})). For the definition of the constituent quark masses see
 (\ref{sigma_shift}) and (\ref{Lbar}) below.

In addition to the one-loop corrections to the constant $G$ at
four-quark vertices, we   modified the current quark masses $m_a^0$
(see (\ref{twoloopcorrect1}) and (\ref{twoloopcorrect2})).  This is to
avoid the problem of double counting of the 't Hoot contribution in
gap equations which was encountered by the author in
\cite{Kleva_92}. After the redefinition of the constant $G$ and of the
current quark masses, we can guarantee that in the large-$N_c$ limit
the mass spectrum of mesons and the gap equations, derived from the
new Lagrangian with modified four-quark vertices and current quark masses,
are the same as those
obtained from the original Lagrangian with six-quark vertices.

Now we can bosonize  Lagrangian (\ref{LGus}).
By introducing auxiliary scalar $\sigma$ and pseudoscalar $\phi$
fields, we obtain
\cite{Volk_86,Volk_82,Cimen_99}
\ba
{\mathcal L}(\sigma,\phi) =
-\frac12\sum_{a,b=1}^9\left(
\sigma_a(G^{(-)})^{-1}_{ab}\sigma_b + \phi_a(G^{(+)})^{-1}_{ab}\phi_b
\right)-  \nonumber \\
\quad-i~{\rm Tr}\ln \left\{ i{\hat \partial} - \bar m^0+
\sum_{a=1}^9\tau_a(\sigma_a + i\gamma_5 \phi_a) \right\}.
\ea
As we expect, the chiral symmetry is spontaneously broken due
to strong attraction of quarks in the scalar channel and
the scalar isoscalar fields acquire nonzero vacuum expectation
values $\langle\sigma_a\rangle_0\not=0$ $(a=8,9)$.
These values are related to basic model parameters
$G$, $m^0$, and $\Lambda$ via gap equations as
it will be shown in the next Section.
Therefore, we first have to shift the $\sigma$ fields by a proper value
so that the new fields  have zero vacuum expectation values:
\beq
\sigma_a=\sigma_a'-\mu_a+\bar\mu_a^0,\qquad \langle\sigma_a'\rangle_0=0,
\label{sigma_shift}
\eeq
where $\mu_a=0, \quad (a=1,\dots ,7)$,
 $\mu_8=m_u$, $\mu_9=-m_s/\sqrt{2}$ and
$\bar\mu_a^0=0, \quad (a=1,\dots ,7)$,
 $\bar\mu^0_8=\bar m_u^0$, $\bar \mu^0_9=-\bar m_s^0/\sqrt{2}$.
After this shift we obtain:
\beq
{\mathcal L}(\sigma',\phi) =
\quad L_G(\sigma',\phi)
-i~{\rm Tr}\ln \left\{ i{\hat \partial} -  m+
\sum_{a=1}^9\tau_a(\sigma_a' + i\gamma_5 \phi_a) \right\},
\label{Lbar}
\eeq
where
\ba
&&L_G(\sigma',\phi)=-\frac12\sum_{a,b=1}^9
(\sigma_a'-\mu_a+\bar\mu_a^0)\left(G^{(-)}\right)^{-1}_{ab}
(\sigma_b'-\mu_a+\bar\mu_a^0)-\nonumber\\
&&-\frac12 \sum_{a,b=1}^9
\phi_a\left(G^{(+)}\right)^{-1}_{ab}
\phi_b,  \label{LG}
\ea
and $m$ is a diagonal matrix of constituent quark masses for
different flavors.
From Lagrangian (\ref{Lbar}) we take only those terms  (in momentum space)
which are linear, squared, cubic, and quadruple
in scalar and pseudoscalar
fields.\footnote{Despite that the scalar fields
are of the main interest in                                    
this paper, we still need  pseudoscalar fields to fix
the model parameters.}
\ba
{\mathcal L}(\sigma',\phi)&=&L_G(\sigma',\phi)+
\tr\Biggl[I_2^{\Lambda}(m)
((\partial_\mu\sigma')^2+(\partial_\mu\phi)^2)-4 m I^\Lambda_1(m)\sigma'+\nonumber\\
&+&2I_1^{\Lambda}(m)(\sigma'^2+\phi^2)-
4m^2I_2^{\Lambda}(m)\sigma'^2+\nonumber\\
&+&4mI^{\Lambda}_2(m)\sigma'(\sigma'^2+\phi^2)^2-
I^{\Lambda}_2(m)(\sigma'^2+\phi^2)^2+\nonumber\\
&+&I^{\Lambda}_2(m)[\sigma'-m,\phi]_{-}^2\Biggr],
\label{Lagr:bosonized}\\
&& \sigma'=\sum_{a=1}^9\sigma_a\tau_a,\qquad \phi=\sum_{a=1}^9\phi_a\tau_a,
\ea
where  ``tr''
means calculating the trace over  $\tau$-matrix expressions
and $[\dots]_-$ stands for a commutator.
(The calculation of ``tr'' is explained in details in \cite{Volk_86})
The expression for $I^\Lambda_1(m_a)$  in
Euclidean metric is given in (\ref{I1}).
The integrals $I^{\Lambda}_2(m_a)$ are also calculated
in Euclidean space-time
\beq
I^{\Lambda}_2(m_a)=\frac{N_c}{16\pi^2}
\left(\ln\left({\Lambda^2\over m_a^2}+1\right)-
{\Lambda^2\over \Lambda^2+m_a^2}\right).
\label{I2}
\eeq
Then, we  renormalize the fields in (\ref{Lagr:bosonized}) so that
the kinetic terms of the effective Lagrangian are of the conventional
form, and diagonalize the isoscalar sector.
\ba
&& \bar{\mathcal L}(\sigma^r,\phi^r)=
\bar L_G(\sigma^r,\phi^r)+\nonumber\\
&&\qquad+
\tr\Biggl[\frac{1}{4}((\partial_\mu\sigma^r)^{2}+(\partial_\mu\phi^r)^2)-
4mg I_1^{\Lambda}(m)\sigma^r+2g^2 I^{\Lambda}_1(m)(\sigma^{r\; 2}+
Z\phi^{r\;2})+\nonumber\\
&&\qquad+\frac14[m,\phi^r]_{-}^2-
 m^2 \sigma^{r\; 2} +
 m g\sigma^{r}(\sigma^{r\; 2}+Z\phi^{r\;2})-
\frac{g}{2}[m,\phi^{r}]_{-}[\sigma^{r},\phi^{r}]_{-}-\nonumber\\
&&\qquad-{g^2\over 4}((\sigma^{r\; 2}+Z\phi^{r\; 2})^2 -
[\sigma^{r},\phi^{r}]_{-}^2
)\Biggr],
\label{Lagr:bosonized:r}\\
&&\qquad \sigma^{r}=\sum_{a=1}^9\sigma_a^{r}\tau_a,
\qquad \phi^r=\sum_{a=1}^9\phi_a^{r}\tau_a.
\ea
For $\bar L_G$ we have:
\ba
&&\bar L_G(\sigma^r,\phi^r)=-\frac12\sum_{a,b=1}^9
(g_a\sigma^r_a-\mu_a+\bar\mu_a^0)\left(G^{(-)}\right)^{-1}_{ab}
(g_b\sigma^r_b-\mu_b+\bar\mu_b^0)-\nonumber\\
&&\quad-
\frac{Z}{2}\sum_{a,b=1}^9
g_a\phi^r_a\left(G^{(+)}\right)^{-1}_{ab}
g_b\phi^r_b.  \label{LG1}
\ea
Here we introduced
Yukawa coupling  constants $g_a$:
 \beq
\sigma'_a=g_a\sigma^r_a,\qquad \phi_a=\sqrt{Z}g_a\phi^r_a, \label{renorm}
\eeq
\ba
&& g_1^2=g_2^2=g_3^2=g_8^2=g_u^2=[4I^\Lambda_2(m_u)]^{-1},\nonumber\\
&&g_4^2=g_5^2=g_6^2=g_7^2=[4I^\Lambda_2(m_u,m_s)]^{-1}, \nonumber \\
&&\quad g_9^2=g_s^2=[4I^\Lambda_2(m_s)]^{-1}, \\
&&I^\Lambda_2(m_u,m_s)=
{N_c\over (2\pi)^4}\int d^4_e k {\theta (\Lambda^2 -k^2)
\over (k^2 + m^2_u)(k^2 + m^2_s)}=\nonumber\\
&&={3\over (4\pi)^2(m_s^2-m_u^2)}\left[m_s^2\ln\left({\Lambda^2
\over m_s^2}+1 \right) - m_u^2\ln\left({\Lambda^2\over m_u^2}+1
\right) \right],
\label{ga_0}\\
&& Z=\left(1-\frac{6m_u}{M_{A_1}^2}\right)^{-1}\approx 1.44 \label{Z} ,
\ea
where we have taken into account $\pi$-$A_1$-transitions
leading to an additional $Z$ factor, with $M_{A_1}$ being
the mass of axial-vector meson (see \cite{Volk_86}).
The renormalized scalar and pseudoscalar
fields in (\ref{Lagr:bosonized:r})--(\ref{renorm}) are marked with the superscript $r$.

The mass formulae for isovectors and isodublets follow immediately from
(\ref{Lagr:bosonized:r}). One just has to look up  the coefficients at
$\sigma^{r\; 2}$ and $\phi^{r\; 2}$. There are still nondiagonal
terms in (\ref{LG1}) in the isoscalar sector. This problem is solved by
choosing the proper mixing angles both for the scalars and pseudoscalars
(see, e.~g., \cite{Cimen_99}).
As we are going to introduce the glueball field, the mixing with
scalar isoscalar quarkonia will change the situation. One has to
consider the mixing among three states, which cannot be described
by a single angle.
For simplicity, in our estimations we
resort to a numerical diagonalization procedure, not to the algebraic one.
Concerning the pseudoscalar sector, one can avail oneself with
the results given in \cite{Cimen_99}.
All what concerns dealing with the glueball is discussed in
the next Section.

\section{Nambu--Jona-Lasinio model with dilaton}

As we have already mentioned above, we introduce the glueball field
into our effective Lagrangian obtained in the previous
Section, as a dilaton. For this purpose,  we use the following
principle. Insofar as the QCD Lagrangian is scale-invariant
in the chiral limit, we suppose that our effective meson
Lagrangian, motivated by QCD, has also to be  scale-invariant
both before and  after SBCS in the case when the current quark
masses  are equal to zero. As a result, we come to
the following prescription: the dimensional model parameters
$G$, $\Lambda$, $K$, and $m_a$ are replaced by the following rule:
$G\to G(\chi_c/\chi)^2$, $K\to K(\chi_c/\chi)^5$,
$\Lambda\to\Lambda (\chi/\chi_c)^2$, $m_a\to m_a(\chi/\chi_c)$,
where $\chi$ is the dilaton field with the vacuum expectation
value $\chi_c$. But there are terms that break scale invariance.
They are the terms containing current quark masses; the scale anomaly of QCD,
reproduced by the dilaton potential; and terms of the type
$\hps\phi_0^2$ and $\hsc\sigma_0^2$ (see (\ref{anomalynotscaled})),
induced by gluon
anomalies in the meson Lagrangian.

As it was mentioned in the previous paragraph,
the current quark masses break scale invariance and,
therefore, should not be multiplied by dilaton fields.
The modified current quark masses $\bar m^0$ are also not
multiplied by dilaton fields.
In particular, this transforms formula (\ref{sigma_shift})
to what follows:
\beq
	\sigma_a=\sigma_a'-\mu_a{\chi\over\chi_c}+\bar\mu_a^0.
\eeq
Finally, we come to the following Lagrangian:
\beq
\bar{\mathcal L}(\sigma^r,\phi^r,\chi)={\mathcal L}(\chi)+
L_{\mathrm kin}(\sigma^r,\phi^r)+
\bar L_G(\sigma^r,\phi^r,\chi)+L_{\rm 1-loop}
+\Delta L_{\mathrm an}.
\label{Lagr:bosonized:chi}
\eeq
Here ${\mathcal L}(\chi)$ is the pure dilaton Lagrangian
\beq
{\mathcal L}(\chi)=\frac12(\partial_\nu\chi)^2-V(\chi)
\eeq
with the  potential
\ba
V({\chi})=B\left({\chi\over {\chi}_0} \right)^4\left[ \ln \left({\chi\over
{\chi}_0} \right)^4 -1 \right] \label{chi}
\ea
that has a
minimum at $\chi = \chi_0$, and the parameter $B$ representing the vacuum
energy when there are no quarks. The curvature of the potential at its
minimum determines the bare glueball mass
\ba m_g = {4\sqrt{B}\over {\chi}_0}.
\label{Defm_g}
\ea
The part $L_{\mathrm kin}(\sigma^r,\phi^r)$ of Lagrangian (\ref{Lagr:bosonized:chi})
contains pure kinetic terms:
\beq
L_{\mathrm kin}(\sigma^r,\phi^r)=
\frac12\sum_{a=1}^9\left(\left(\partial_\nu\sigma_a^r\right)^2+
\left(\partial_\nu\phi_a^r\right)^2\right).
\eeq
The next term reads
\ba
&&\bar L_G(\sigma^r,\phi^r,\chi)=\nonumber\\
&&\qquad=-\frac12\left({\chi\over\chi_c}\right)^2\sum_{a,b=1}^9
\left(g_a\sigma_a^r-\mu_a{\chi\over\chi_c}+
\bar\mu^0_a\right)\left(G^{(-)}\right)^{-1}_{ab}\times\nonumber\\
&&\qquad\times
	\left(g_b\sigma_b^r-\mu_b{\chi\over\chi_c}
+\bar\mu^0_b\right)-\nonumber\\
&&\qquad -\frac{Z}{2}\left({\chi\over\chi_c}\right)^2\sum_{a,b=1}^9
g_a\phi_a^r\left(G^{(+)}\right)^{-1}_{ab}g_b\phi_b^r.
\label{LGr}
\ea

The sum of one-loop quark diagrams is denoted as $L_{\rm 1-loop}$:
\ba
&&L_{\rm 1-loop}=\tr\Biggl[-4m g I^\Lambda_1(m)\sigma^r
\left({\chi\over\chi_c}\right)^3
+ 2g^2I^\Lambda_1(m)(\sigma^{r\; 2}+\phi^{r\;2})
\left( {\chi\over\chi_c}\right)^2-\nonumber\\
&&\quad-m^2g^2\sigma^{r\; 2} \left({\chi\over\chi_c}\right)^2
+m g {\chi\over\chi_c}\sigma^r(\sigma^{r\; 2}+\phi^{r\; 2})-\nonumber\\
&&\quad-{g^2\over 4}(\sigma^{r\; 2}+\phi^{r\; 2})^2\Biggr].
\ea

As one can see, expanding (\ref{LGr})
in a power series of $\chi$, we can extract a term that is of order $\chi^4$.
It can be absorbed by the term in the pure dilaton potential
which has the same degree of $\chi$.
Obviously, this leads only
to a redefinition of constants $B$ and $\chi_0$ which anyway
are not known from the very beginning.
Moreover, saying in advance, the terms like $\chi^4$ do not contribute to the
divergence of the dilatation current (\ref{Ward}) because of
their scale invariance.

If  the procedure of the scale invariance restoration
of this Lagrangian is implemented,
the part induced by gluon anomalies
also becomes scale-invariant. To avoid this, one should
subtract this part in the scale-invariant form and add
it in a scale-breaking (SB) form. This is achieved by
including the  term $\Delta L_{\mathrm an}$:
\beq
\Delta L_{\mathrm an}=-L_{\mathrm an}\left(\frac{\chi}{\chi_c}\right)^2
+L_{\mathrm an}^{\mathrm SB}.
\eeq
The term $L_{\mathrm an}$ was introduced in (\ref{anomalynotscaled}).
In $L_{\mathrm an}$, we will use renormalized $\sigma^r_0$ and $\phi^r$ fields
in place of $\sigma_0$ and $\phi_0$, however, taking into account
the effects of nonzero vacuum expectation
value of $\sigma_0$.
Let us define the scale-breaking term $L_{\mathrm an}^{\mathrm SB}$.
The coefficients $\hsc$ and $\hps$ in (\ref{anomalynotscaled})
can be determined by
comparing them with the  terms in (\ref{LGr}) that describe the singlet-octet mixing.
We obtain
\beq
\hps=-\frac{3}{2\sqrt{2}} g_u g_s Z (G^{(+)})^{-1}_{89},\qquad
\hsc=\frac{3}{2\sqrt{2}} g_u g_s  (G^{(-)})^{-1}_{89}.
\eeq
If  these terms were to be made scale invariant,
one should insert $(\chi/\chi_c)^2$ into them. However,
as the gluon anomalies break scale invariance, we
introduce the dilaton field into these terms in a more complicated way.
The inverse matrix  elements $(G^{(+)})^{-1}_{ab}$ and
$(G^{(-)})^{-1}_{ab}$,
\beq
(G^{(+)})^{-1}_{89}={-4\sqrt{2}m_u K I_1^\Lambda(m_u)\over
G^{(+)}_{88}G^{(+)}_{99}-(G^{(+)}_{89})^2},
\eeq
\beq
(G^{(-)})^{-1}_{89}={4\sqrt{2}m_u K I_1^\Lambda(m_u)\over
G^{(-)}_{88}G^{(-)}_{99}-(G^{(-)}_{89})^2},
\eeq
are determined by two different interactions.
The numerators are fully defined by the 't Hooft
interaction that leads to anomalous terms (\ref{anomalynotscaled})
breaking scale invariance, therefore we do not introduce
here dilaton fields. The denominators are determined by
constant $G$ describing the main four-quark interaction, and the
dilaton field is inserted into it, according to the prescription
given in the beginning of this Section.
Finally, we come to the following form of $L_{\mathrm an}^{\mathrm SB}$:
\ba
&&L_{\mathrm an}^{\mathrm SB}=\left(-\hps\phi_0^{r\;2}+
\hsc\left(\sigma_0^r-F_0\frac{\chi}{\chi_c}+F_0^0\right)^2\right)\left(
\frac{\chi}{\chi_c}\right)^4,\\
&& F_0=
\frac{\sqrt{2}m_u}{\sqrt{3}g_u}+\frac{m_s}{\sqrt{6}g_s},\qquad
F_0^0=\frac{\sqrt{2}\bar m_u^0}{\sqrt{3}g_u}+\frac{\bar m_s^0}{\sqrt{6}g_s}.
\ea
From it, we immediately obtain the term $\Delta L_{\mathrm an}$:
\beq
\Delta L_{\mathrm an}
=\left(\hps\phi_0^{r\; 2}-\hsc\left(\sigma_0^r-F_0\frac{\chi}{\chi_c}+F_0^0
\right)^2\right)\left(
\frac{\chi}{\chi_c}\right)^2\left(1-\left(\frac{\chi}{\chi_c}\right)^2\right)
\label{Deltaanomalynotscaled}.
\eeq

Let us now consider the vacuum expectation value of the
divergence of the dilatation current calculated from the
potential of the effective  meson-dilaton Lagrangian:
\ba
 &&\langle\partial_{\mu}S^{\mu}\rangle=\left(\sum_{a=8}^9
\sigma_a^r{\partial V\over\partial \sigma_a^r}+
 \chi{\partial V\over\partial \chi}-4V\right)
\Biggr\vert_{\begin{array}{l}\scriptstyle \chi = \chi_c\, \\[-2mm]
\scriptstyle\sigma_a^r = 0
\end{array}}=\nonumber\\
&&\quad=4B\left({\chi_c\over\chi_0}\right)^4
-2\hsc\left(F_0-F_0^0\right)^2
-\sum_{q=u,d,s}\bar m^0_q\langle\bar qq\rangle.
\label{dilaton:current}
\ea
Here $V=V(\chi)+\bar V(\sigma^r,\phi^r,\chi)$,
and $\bar V(\sigma^r,\phi^r,\chi)$
is the potential part of Lagrangian $\bar{\mathcal L}(\sigma^r,\phi^r,\chi)$
that does not contain the pure dilaton potential.
The expression given in  (\ref{dilaton:current}) is simplified by using
 the following relation  of the quark condensates
 to integrals $I_1^\Lambda(m_u)$ and $I_1^\Lambda(m_s)$:
\beq
4m_q I^\Lambda_1(m_q)=-\langle\bar qq\rangle_0,
\qquad (q=u,d,s), \label{I1toQQ}
\eeq
and that these integrals are connected with constants $G^{(-)}_{ab}$
through gap equations, as it will be shown in the next Section
(see (\ref{gapeqbegin}) and (\ref{gapeq2}) below).
Comparing the QCD expression (\ref{Ward})
with (\ref{dilaton:current}), one can see that
the term $\sum m^0_q\langle\bar qq\rangle$
on the left-hand side is canceled by the corresponding
contribution on the right-hand side.
Equating the right hand sides of (\ref{Ward}) and (\ref{dilaton:current}),
\ba
&&\Cg-\sum_{q=u,d,s}m^0_q\langle\bar qq\rangle=\nonumber\\
&&\quad=4B\left({\chi_c\over\chi_0}\right)^4
-2\hsc\left(F_0-F_0^0\right)^2-\sum_{q=u,d,s}\bar m^0_q\langle\bar qq\rangle,
\ea
we obtain the correspondence
\beq
\Cg= 4B\left(\chi_c\over\chi_0\right)^4+
\sum_{a,b=8}^9(\bar\mu_a^0-\mu_a^0)(G^{(-)})_{ab}^{-1}(\mu_b-\bar\mu_b^0)
-2\hsc\left(F_0-F_0^0\right)^2, \label{Cg}
\eeq
where $\mu_a^0=0\quad (a=1,\dots 7)$, $\mu_8^0=m^0_u$,
and $\mu_9^0=-m_s/\sqrt{2}$.
This equation relates the gluon condensate, whose value we take
from other models (see, e.~g., \cite{Narison96}),
to the model parameter $B$. The next step is to investigate
the gap equations.

\section{Equations}

As usual,
gap equations follow from the requirement that
the terms linear in $\sigma^r$ and $\chi'$ should be absent in our Lagrangian:
\ba
&&{{\delta}\bar{\mathcal L}\over {\delta}\sigma_8^r}
\biggr\vert_{(\phi^r,\sigma^r,{\chi'}) = 0}=0
,\quad
{{\delta}\bar{\mathcal L}\over {\delta}\sigma_9^r}
\biggr\vert_{(\phi^r,\sigma^r,{\chi'}) = 0}=0
,\quad
{{\delta}\bar{\mathcal L}\over {\delta}\chi}
\biggr\vert_{(\phi^r,\sigma^r,{\chi'}) = 0}=0.
\label{var}
\ea
Here, the field $\chi'=\chi-\chi_c$ with
a zero vacuum expectation value $\langle\chi'\rangle_0=0$,
is associated with the glueball field. In further calculations,
the Lagrangian is expanded in power series of $\chi'$.
As a result the following equations are obtained:
\ba
(m_u-\bar{m}_u^0)(G^{(-)})^{-1}_{88} - {{m_s-\bar{m}_s^0}\over \sqrt2}(G^{(-)})^{-1}_{89} -
8m_uI^\Lambda_1(m_u)& =& 0 , \label{gapeqbegin} \\
(m_s-\bar{m}_s^0)(G^{(-)})^{-1}_{99} - {\sqrt2}(m_u-\bar{m}_u^0)(G^{(-)})^{-1}_{98} -
8 m_sI^\Lambda_1(m_s) &=& 0 ,\label{gapeq2} \\
4B\left({\chi_c\over {\chi}_0} \right)^3{1\over \chi_0}
\ln \left({\chi_c\over {\chi}_0} \right)^4 +
{1\over \chi_c}\left(\sum_{a,b=8}^9\bar\mu_a^0(G^{(-)})_{ab}^{-1}(\bar\mu_b^0-3\mu_b)\right)-&&\nonumber\\
-\frac{2\hsc}{\chi_c}\left(F_0-F_0^0\right)^2 &=& 0.
\label{Gapeqs}
\ea

Using
(\ref{twoloopcorrect1}) and (\ref{twoloopcorrect2}), one can rewrite
equations (\ref{gapeqbegin}) and (\ref{gapeq2})  in
the well-known form \cite{Kleva_92}:
\ba
m_u^0&=&m_u-8 G m_u I_1^{\Lambda}(m_u)-32K m_u m_s I_1^\Lambda(m_u)I_1^\Lambda(m_s),
\label{gap:eq:u}\\
m_s^0&=&m_s-8 G m_s I_1^{\Lambda}(m_s)-32K(m_u I_1^\Lambda(m_u))^2.
\label{gap:eq:s}
\ea

To define the masses of quarkonia and the glueball, let
us consider the part of Lagrangian (\ref{Lagr:bosonized:chi})
which is quadratic in fields $\sigma^r$ and $\chi'$
and which we denote as $L^{(2)}$
\ba
L^{(2)}(\sigma,\phi,\chi') &=& -{1\over 2}g_8^2\{[(G^{(-)})^{-1}_{88}
-8I^\Lambda_1(m_u)] + 4m^2_u\}{\sigma^r}^2_8-  \nonumber \\
&-&{1\over 2}g_9^2\{[(G^{(-)})^{-1}_{99} -8I^\Lambda_1(m_s)] + 4m^2_s\}{\sigma^r}^2_9-
  \nonumber \\
&-&g_8g_9(G^{(-)})^{-1}_{89}{\sigma}_8^r{\sigma}_9^r
-\frac{M_g^2}{2}{\chi'}^2+\nonumber\\
&+&\!\!\!\sum_{a,b=8}^9\frac{\bar\mu_a^0}{\chi_c}(G^{(-)})^{-1}_{ab}
g_b\sigma_b^r\chi'+\frac{4\hsc (F_0-F_0^0)}{\chi_c\sqrt{3}}
\left({\sigma_9^r}-\sigma_8^r\sqrt{2}\right)\chi'
\label{BAchi},
\ea
where
\ba
&&M_g^2=\chi_c^{-2}(4\Cg+\sum_{a,b=8}^{9}
\bar\mu^0_a (G^{(-)})^{-1}_{ab}(2\bar\mu^0_b-\mu_b)+\nonumber\\
&&\qquad+\sum_{a,b=8}^{9}4\mu^0_a (G^{(-)})^{-1}_{ab}(\mu_b-\bar\mu_b^0)
-\hsc4 F_0^2+ 4\hsc (F_0^0)^2)
\ea
is the glueball mass before taking account of mixing effects.

From this Lagrangian, after diagonalization, we obtain
the masses of three scalar meson states: $\sigma_{\rm I}$, $\sigma_{\rm II}$,
and $\sigma_{\rm III}$, and a matrix of mixing coefficients $b$ that
connects the nondiagonalized fields $\sigma_8\equiv\sigma_u,
\sigma_9\equiv\sigma_s,\chi'$
with the physical ones $\sigma_{\mathrm I},
\sigma_{\mathrm II},\sigma_{\mathrm III}$
\beq
\left(
\begin{array}{c}
\sigma_u\\ \sigma_s \\ \chi'
\end{array}
\right)=
\left(
\begin{array}{lll}
b_{\sigma_u\sigmaI} & b_{\sigma_u\sigmaII}&b_{\sigma_u\sigmaIII}\\
b_{\sigma_s\sigmaI} & b_{\sigma_s\sigmaII}&b_{\sigma_s\sigmaIII}\\
b_{\chi'\sigmaI} & b_{\chi'\sigmaII}&b_{\chi'\sigmaIII}\\
\end{array}
\right)\left(
\begin{array}{c}
\sigmaI\\ \sigmaII \\ \sigmaIII
\end{array}
\right).
\eeq

\section{Model parameters and numerical estimates}

The basic parameters of our model are $G$, $K$, $\Lambda$, $m_u$,
and $m_s$. After the dilaton fields are introduced, they keep their
values \cite{Volk_86,Cimen_99}:
\ba
&&m_u=280\;\mbox{MeV},\;m_s=420\;\mbox{MeV},\;\Lambda =1.26\;\mbox{
GeV},\nonumber\\
&&G=4.38\;\mbox{GeV}^{-2},\;K=11.2\;\mbox{GeV}^{-5}. \label{paramet}
\ea
Moreover, new three parameters
$\chi_0$, $\chi_c$, and $B$ appear. To fix the new parameters,
one should use equations (\ref{Cg}), (\ref{Gapeqs}), and
the physical glueball mass.
As a result we  obtain for $\chi_0$ and $B$:
\beq
\chi_0=\chi_c \exp \left(-{
\sum_{a,b=8}^{9}
\bar\mu^0_a (G^{(-)})^{-1}_{ab}(3\mu_b-\bar\mu^0_b)+
2\hsc\left(F_0-F_0^0\right)^2
 \over 4[\Cg-(\bar\mu^0_a-\mu^0_a) (G^{(-)})^{-1}_{ab}(\mu_b-\bar\mu^0_b)+2\hsc\left(F_0-F_0^0\right)^2]}\right),
\eeq
\beq
B=\frac{\Cg-(\bar\mu^0_a-\mu^0_a) (G^{(-)})^{-1}_{ab}(\mu_b-\bar\mu^0_b)+2\hsc\left(F_0-F_0^0\right)^2}{4}
\left(\frac{\chi_0}{\chi_c}\right)^4.
\eeq
We adjust the parameter $\chi_c$ so that
the mass of the heaviest scalar meson, $\sigmaIII$, would be
either 1500 MeV  or 1710 MeV.
The result of our fit for both cases is given in
Table~\ref{T:spectr}. One also will find
the mixing coefficients in Table~\ref{mix}.

\section{Decay widths}

Once all parameters are fixed, we can  estimate
the decay widths for the main strong decay
modes of scalar mesons: $\sigma_l\to\pi\pi$,$KK$,
$\eta\eta$, $\eta\eta'$, and  $4\pi$
where $l=\mathrm I,II,III$.

Note  that, in the energy region under consideration ($\sim 1500$ MeV),
we work on the brim of the validity of exploiting the chiral symmetry
that was used to construct our effective Lagrangian.
Thus, we can consider our results as rather qualitative.

Let us start with the lightest scalar isoscalar meson state $\sigma_{\mathrm I}$,
associated with $f_0(400-1200)$. This state decays into pions.
This is the only strong decay mode, because $\sigma_{\mathrm I}$  is too light
for other channels to be open.
The amplitude describing its decay into pions
has the form:
\beq
A_{\sigma_{\mathrm I}\to \pi^+\pi^-} =
2A_{\pi^+\pi^-}^g b_{\chi'\sigma_{\mathrm I}}
 +2A^u b_{\sigma_u\sigma_{\mathrm I}},
\eeq
\beq
A_{\pi^+\pi^-}^g=
  -\frac{M_\pi^2}{\chi_c},\quad A^u=2 g_u m_u Z,
\eeq
where $A_{\pi^+\pi^-}^g$ is the contribution from the glueball
component; and $A^u$, from the ($\bar u u$) quarkonium one.
The coefficients $b_{\chi'\sigma_{\mathrm I}}$ and $b_{\sigma_u\sigma_{\mathrm I}}$
represent the  corresponding elements of the
$3\times 3$ mixing matrix for scalar isoscalar states (see Table~\ref{mix}).
Both contributions have equal signs and add to the
the width of $\sigma_{\rm I}$.

To calculate the decay width of a meson into two mesons, one can use
the following formula:
\beq
\Gamma=\frac{|A|^2}{16\pi M^3}\frac{\lambda^{1/2}(M^2,M_1^2,M_2^2)}{r}, \label{width}
\eeq
where $A$ is the amplitude of the process; $M$ is the mass of a decaying particle;
$M_1$ and $M_2$ are masses of secondary particles; $r$ is the dimension of
the permutation symmetry group
in the phase space of final states. The function $\lambda(x,y,z)$
is defined as follows \cite{Kajantie}:
\beq
\lambda(x,y,z)=(x-y-z)^2-4yz.
\eeq
For the decay of $\sigma_{\mathrm I}$ into pions, formula (\ref{width}) can be
rewritten in a simpler form
\beq
\Gamma_{\sigma_{\mathrm I}\to\pi^+\pi^-}=
\frac{|A_{\sigma_{\mathrm I}\to \pi^+\pi^-}|^2}{16\pi M_{\sigma_{\mathrm I}}}
\sqrt{1-\frac{4 M_\pi^2}{M_{\sigma_{\mathrm I}}^2}}.
\eeq
Using isotopic symmetry, we obtain the total width
\beq
\Gamma_{\sigma_{\mathrm I}\to\pi\pi}=
\frac32\Gamma_{\sigma_{\mathrm I}\to\pi^+\pi^-}\approx
820 \mbox{ MeV} \label{sigmawidth}
\eeq
for the case when the model parameters are
fixed for the state $\sigma_{\mathrm III}$ identified with $f_0(1500)$,
and
\beq
\Gamma_{\sigma_{\mathrm I}\to\pi\pi}\approx
830 \mbox{ MeV}.\label{sigmawidth2}
\eeq
for the case $\sigma_{\mathrm III}\equiv f_0(1710)$.
The experimental value is known with a large uncertainty and is
reported to lie in the interval from 600  to 1000 MeV \cite{PDG}.

The amplitude describing the decay of the state $\sigma_{\mathrm II}$ which we identify
with $f_0(980)$ into pions also consists of two parts
\beq A_{\sigma_{\mathrm II}\to
\pi^+\pi^-} =
2A_{\pi^+\pi^-}^g b_{\chi'\sigma_{\rm II}}
+2A^u b_{\sigma_u\sigma_{\rm II}}.
\eeq
Here the glueball contribution is small again and the
quarkonium determines the decay width, however,
in this case both contributions are opposite in sign and
slightly compensate each other.
The width of the state $\sigma_{\rm II}$ is close to that
obtained in the model without glueballs \cite{Cimen_99}.
We obtain
\beq
\Gamma_{\sigma_{\mathrm II}\to\pi\pi}\approx 28
\mbox{ MeV},
\eeq
if $\sigma_{\mathrm III}\equiv f_0(1500)$ and
\beq
\Gamma_{\sigma_{\mathrm II}\to\pi\pi}\approx 26 \mbox{ MeV},
\eeq
if $\sigma_{\mathrm III}\equiv
f_0(1710)$.
The experiment gives for the decay of $\sigma_{\mathrm II}$
into pions a value lying within the
range 30 -- 70 MeV \cite{PDG}.

Now let us proceed with decays of $\sigma_{\mathrm III}$.
The process
$\sigma_{\mathrm III}\to\pi^+\pi^-$ is given by the amplitude
\beq
A_{\sigma_{\mathrm III}\to\pi^+\pi^-}=
2A_{\pi^+\pi^-}^{g}b_{\chi'\sigma_{\rm III}}+
2A^{u}b_{\sigma_u\sigma_{\rm III}}
\eeq
that consists of
two parts. The first part represents the contribution
from the pure glueball.
This contribution is small (since it is
proportional to the pion mass squared),
and the process is determined by
the second part that describes the decay of the
quarkonium component.
As a result, the width of the decay $\sigma_{\mathrm III}\to\pi\pi$
if $\sigma_{\mathrm III}\equiv f_0(1500)$ is
\beq
\Gamma_{\sigma_{\mathrm III}\to\pi\pi}=
\frac{3}{2}\Gamma_{\sigma_{\mathrm III}\to\pi^+\pi^-}\approx 14\; \mbox{MeV},
\eeq
and, if $\sigma_{\mathrm III}\equiv f_0(1710)$,
\beq
\Gamma_{\sigma_{\mathrm III}\to\pi\pi}\approx 8\; \mbox{MeV}.
\eeq

In the case of $K\bar K$ channels, the contribution
of the pure glueball is also proportional to the kaon
mass squared, and is rather large as compared to the pions case.
The amplitude of the decay $\sigma_{\mathrm III}\to K^+K^-$
consists of  three parts
\beq
A_{\sigma_{\mathrm III}\to K^+ K^-}=
A_{KK}^{g}b_{\chi'\sigma_{\mathrm III}}+
A_{KK}^{u}b_{\sigma_u\sigma_{\mathrm III}}
+A_{KK}^{s}b_{\sigma_s\sigma_{\mathrm III}},
\eeq
where the pure glueball decay into $K^+ K^-$ is
given by the amplitude
\beq
A_{KK}^{g}=
-\frac{2 M_K^2}{\chi_c}.
\eeq
The quarkonium contributions are
\ba
A_{KK}^{u}&=&
2g_u Z\left(\frac{m_u+m_s}{2}\left(\frac{F_\pi}{F_K}\right)^2+
\frac{m_s(m_u-m_s)}{m_u+m_s}\right),\\
A_{KK}^{s}&=&-
4\sqrt{2} g_s Z\left(\frac{m_u+m_s}{2}\left(\frac{F_s}{F_K}\right)^2+
\frac{m_u(m_s-m_u)}{m_u+m_s}\right),
\ea
where $F_\pi$ and $F_K$ are the pion and kaon weak decay
constants, respectively, and  $F_s=m_s/(g_s\sqrt{Z})$.
In the case when $\sigma_{\mathrm III}$ is $f_0(1500)$, we have
\beq
 \Gamma_{\sigma_{\mathrm III}\to K\bar K}=
 \Gamma_{\sigma_{\mathrm III}\to K^+K^-}+
\Gamma_{\sigma_{\mathrm III}\to K^0\bar K^0}=
2\Gamma_{\sigma_{\mathrm III}\to K^+K^-}\approx
29\; \mbox{MeV},
\eeq
and in the other case ($\sigma_{\mathrm III}\equiv f_0(1710)$)
\beq
 \Gamma_{\sigma_{\mathrm III}\to K\bar K}
\approx 60\;\mbox{MeV}.
\eeq

The amplitude of the decay of $\sigma_{\mathrm III}$
into $\eta\eta$ and $\eta\eta'$ can also be
considered in the same manner. The only complication is the singlet-octet
mixing in the pseudoscalar sector and additional vertices coming
from $\Delta L_{\mathrm an}$.
 The corresponding amplitude is
\beq
A_{\sigma_{\mathrm III}\to\eta\eta}=
2A_{\eta\eta}^g b_{\chi'\sigma_{\rm III}}
+2A^u\sin^2\bar\theta b_{\sigma_u \sigma_{\mathrm III}}+
+2A^s\cos^2\bar\theta b_{\sigma_s\sigma_{\mathrm III}}+
2A^{\mathrm an}_\phi\sin^2 \theta b_{\chi'\sigma_{\rm III}},
\eeq
\ba
A_{\eta\eta}^g&=&-\frac{M_\eta^2}{\chi_c},\\
A^{\mathrm an}_\phi&=
&-\frac{2\hps}{\chi_c},
\ea
where $\bar\theta=\theta-\theta_0$, with $\theta$ being the singlet-octet mixing angle
in the pseudoscalar channel, $\theta\approx -19^\circ$ \cite{Cimen_99}, and $\theta_0$ the
ideal mixing angle, $\tan\theta_0=1/\sqrt{2}$.
The decay widths thereby are:
if $\sigma_{\mathrm III}\equiv f_0(1500)$,
\beq
 \Gamma_{\sigma_{\mathrm III}\to \eta\eta}\approx 25\; \mbox{MeV},
\eeq
and if $\sigma_{\mathrm III}\equiv f_0(1710)$,
\beq
\Gamma_{\sigma_{\mathrm III}\to \eta\eta}\approx 43\; \mbox{MeV}.
\eeq
For the decay of $\sigma_{\mathrm III}$ into $\eta\eta'$,
we have the following amplitude
\beq
A_{\sigma_{\mathrm III}\to\eta\eta'}=-
A^u \sin2\bar\theta b_{\sigma_u \sigma_{\mathrm III}} +
A^s \sin2\bar\theta b_{\sigma_s\sigma_{\mathrm III}}-
A^{\mathrm an}_\phi\sin 2\theta b_{\chi'\sigma_{\mathrm III}}.
\eeq

The direct decay of a bare glueball
into $\eta\eta'$ is absent here. This process occurs
only due to the mixing of the glueball and scalar isoscalar quarkonia
and the anomaly contribution.
The decay widths are as follows:
\beq
\Gamma_{\sigma_{\mathrm III}\to \eta\eta'}\sim 10 \; \mbox{MeV},
\eeq
for $\sigma_{\mathrm III}\equiv f_0(1500)$, and
\beq
 \Gamma_{\sigma_{\mathrm III}\to \eta\eta'} \approx 30\; \mbox{MeV}.
\eeq
for $\sigma_{\mathrm III}\equiv f_0(1710)$.
The estimate for the decay  $f_0(1500)$ into
$\eta\eta'$ is very rough, because the decay is allowed only due to
a finite width of the resonance as its mass lies a little bit below the
$\eta\eta'$ threshold. The calculation is made for the mass of $f_0(1500)$ plus
its half-width. For $f_0(1710)$, we have a more reliable estimate, since
its mass is large enough for the decay to be possible.

Up to this moment we considered only decays into a pair of mesons.
For the state $\sigma_{\mathrm III}$, there is a possibility to decay
into 4 pions. This decay can occur through
intermediate  $\sigma$ ($f_0(400-1200)$) resonance.

The decay through the $\sigma$-resonance  can be
represented as two processes: with two resonances
$\sigma_{\mathrm III}\to\sigma\sigma\to 4\pi$ and one resonance
$\sigma_{\mathrm III}\to\sigma 2\pi\to 4\pi$. The vertices determining
these decays follow from Lagrangian (\ref{Lagr:bosonized:chi}).
The decay of a glueball into two $\sigma$ is given by
the amplitude:
\ba
&&A_{\sigma_{\mathrm III}\to\sigma\sigma}\approx
2A_{\sigma\sigma}^g b_{\chi'\sigma_{\mathrm III}}+
3Z^{-1}A^u b_{\sigma_u\sigma_{\mathrm III}}b_{\sigma_u\sigma_{I}}b_{\sigma_u\sigma_{\mathrm I}}+\nonumber\\
&&\quad+2A_{\sigma}^{\mathrm an}b_{\chi'\sigma_{\mathrm III}}b_{\sigma_u\sigma_{\mathrm I}}^2,
\ea
where $A_{\sigma\sigma}^g$ is
the pure glueball amplitude%
\footnote{
To obtain an approximate estimate for the glueball contribution,
we used the mass of $\sigma_u$ state  before diagonalization
(see the term with $\sigma_8^{r\;2}$ in (\ref{BAchi}))
}:
\beq
A_{\sigma\sigma}^g\approx
-\frac{M_{\sigma_u}^2}{\chi_c},
\eeq
and the anomaly amplitude $A_{\sigma\sigma}^{\mathrm an}$
coming from $\Delta L_{\mathrm an}$ is
\beq
A_{\sigma}^{\mathrm an}=
\frac{2\hsc}{3\chi_c}.
\eeq
The total amplitude describing the decay into 4 pions
through two $\sigma$-resonances is
\ba
A_{\sigma_{\mathrm III}\to\sigma\sigma\to 2\pi^+2\pi^-}&=&
2A_{\sigma_{\mathrm III}\to \sigma\sigma}A_{\sigma\to \pi^+\pi^-}^2
(\Delta_\sigma(s_{12})\Delta_\sigma(s_{34})+\nonumber\\
&&+
\Delta(s_{14})\Delta(s_{23})),
\ea
where
the function $\Delta_\sigma(s)$ appears due to
the resonant structure of the processes
\beq
\Delta_\sigma(s)=(s-M_{\sigma_{\rm I}}^2+i M_{\sigma_{\rm I}}
\Gamma_{\sigma_{\rm I}})^{-1},
\eeq
where $\Gamma_\sigma$ is the decay width of the
$\sigma_{\mathrm I}$ resonance (see (\ref{sigmawidth})).
This function depends on an invariant mass squared $s_{ij}$ defined as follows
\beq
s_{ij}=(k_i+k_j)^2, \qquad (i,j=1,\dots, 4). \label{invariantmasses}
\eeq
Here $i$ and $j$ enumerate the momenta $k_i$ of pions $\pi^+(k_1)$, $\pi^-(k_2)$,
$\pi^+(k_3)$, and $\pi^-(k_4)$.

Now let us consider the decay into $4\pi$ through one
$\sigma$-resonance. The amplitude describing this
process is as follows:
\beq
A_{\sigma_{\mathrm III}\to\sigma 2\pi}=
A_{\sigma 2\pi}^g (b_{\sigma_u\sigma_{\rm I}}b_{\chi'\sigma_{\mathrm III}}+b_{\sigma_u\sigma_{\mathrm III}}b_{\chi'\sigma_{\mathrm I}})+
A_{\sigma 2\pi}^u b_{\sigma_u\sigma_{\mathrm III}}b_{\sigma_u\sigma_{\mathrm I}}.
\eeq
The glueball amplitude is
\beq
A_{\sigma 2\pi}^g=\frac{4 m_u g_u Z}{\chi_c}
\eeq
and  the quarkonium:
\beq
A_{\sigma_{\mathrm III}\to\sigma 2\pi}^u=-4 g_u^2 Z.
\eeq
The glueball contribution prevails over the quarkonium one
in magnitude and is opposite in sign.

The amplitude describing the decay $\sigma_{\rm III}\to 2\pi^+2\pi^-$
through one $\sigma$-resonance is:
\ba
A_{\sigma_{\mathrm III}\to\sigma2\pi\to 2\pi^+2\pi^-}&=&
-A_{\sigma_{\mathrm III}\to\sigma 2\pi}A_{\sigma\to\pi^+\pi^-}
(\Delta_\sigma(s_{12})+\Delta_\sigma(s_{34})+\nonumber\\
&&+\Delta_\sigma(s_{14})+\Delta_\sigma(s_{23})),
\ea

The total amplitude of the decay into $2\pi^+2\pi^-$ via
$\sigma$-resonances is obtained as a cumulative contribution
from both one and two intermediate $\sigma$ mesons:
\beq
A_{\sigma_{\mathrm III}\to 2\pi^+2\pi^-}=A_{\sigma_{\mathrm III}\to\sigma\sigma\to 2\pi^+2\pi^-}+
A_{\sigma_{\mathrm III}\to\sigma 2\pi\to 2\pi^+2\pi^-}.
\eeq

The amplitude describing the decay into $2\pi^0\pi^+\pi^-$ has the
following form:
\beq
A_{\sigma_{\mathrm III}\to 2\pi^0\pi^+\pi^-}=
A_{\sigma_{\mathrm III}\to\sigma\sigma\to 2\pi^0\pi^+\pi^-}+
A_{\sigma_{\mathrm III}\to\sigma 2\pi\to 2\pi^0\pi^+\pi^-},
\eeq
where
\ba
A_{\sigma_{\mathrm III}\to\sigma\sigma\to 2\pi^0\pi^+\pi^-}&=&
4A_{\sigma_{\mathrm III}\to 2\sigma}A_{\sigma\to 2\pi^0}A_{\sigma\to\pi^+\pi^-}
\Delta_\sigma(s_{12})\Delta_\sigma(s_{34}),\\
A_{\sigma_{\mathrm III}\to\sigma2\pi\to 2\pi^0\pi^+\pi^-}&=&
-2A_{\sigma_{\mathrm III}\to\sigma 2\pi}A_{\sigma\to2\pi^0}
(\Delta_\sigma(s_{12})+\Delta_\sigma(s_{34})).
\ea
In this case, $k_1$ and $k_2$
are momenta of the two $\pi^0$, and $s_{12}$ is
their invariant mass squared. Indices 3 and 4 stand for $\pi^+$
and $\pi^-$, respectively. The amplitude $A_{\sigma\to2\pi^0}$
is equal to $0.5 A_{\sigma\to\pi^+\pi^-}$

In the case of the decay into $4\pi^0$, we have
\beq
A_{\sigma_{\mathrm III}\to 4\pi^0}=
A_{\sigma_{\mathrm III}\to\sigma\sigma\to 4\pi^0}+
A_{\sigma_{\mathrm III}\to\sigma 2\pi\to 4\pi^0},
\eeq
where
\ba
A_{\sigma_{\mathrm III}\to\sigma\sigma\to 4\pi^0}&=&
4A_{\sigma_{\mathrm III}\to 2\sigma}A_{\sigma\to 2\pi^0}^2
(\Delta_\sigma(s_{12})\Delta_\sigma(s_{34})+\Delta_\sigma(s_{13})\Delta_\sigma(s_{24})+\nonumber\\
&+&\Delta_\sigma(s_{14})\Delta_\sigma(s_{23})),\\
A_{\sigma_{\mathrm III}\to\sigma2\pi\to 2\pi^0\pi^+\pi^-}&=&
-2A_{\sigma_{\mathrm III}\to\sigma 2\pi}A_{\sigma\to2\pi^0}
(\Delta_\sigma(s_{12})+\Delta_\sigma(s_{13})+\Delta_\sigma(s_{14})+\nonumber\\
&+&\Delta_\sigma(s_{23})+\Delta_\sigma(s_{24})+\Delta_\sigma(s_{34})).
\ea

Let us give numerical estimates for these decay modes.
The decay width of the glueball into four particles is calculated using
the prescript given in \cite{Kajantie}
\ba
&&\Gamma_{4\pi}=
\frac{1}{64 (2\pi)^6 r M_{\sigma_{\mathrm III}}^2 }\times \nonumber\\
&&\quad\times\int_{s_{123}^-}^{s_{123}^+}\! d s_{123}
\int_{s_{12}^-}^{s_{12}^+}\! d s_{12}
\int_{s_{34}^-}^{s_{34}^+}\! d s_{34}
\int_{s_{23}^-}^{s_{23}^+}\! d s_{23}
\int_{-1}^{1} \frac{|A_{\sigma_{\mathrm III}\to 4\pi}|^2
d\zeta}{\sqrt{\lambda(s_{123},s_{12},M_{\pi}^2)(1-\zeta^2)}}, \label{4pi}
\ea
where $A_{\sigma_{\mathrm III}\to 4\pi}$ is the amplitude describing one of the
processes discussed above, $M_{\sigma_{\mathrm III}}$ is the
mass of $\sigma_{\mathrm III}$.
The corresponding two-particle invariant masses are defined in (\ref{invariantmasses})
except for $s_{123}$, the invariant mass of three pions
\beq
s_{123}=(k_1+k_2+k_3)^2.
\eeq
The cosine between the plane formed by 3-momenta $\mathbf{k}_1$,
$\mathbf{k}_2$ and the plane formed by $\mathbf{k}_3$, $\mathbf{k}_4$ in
the rest frame of three mesons ($\mathbf{k}_1+\mathbf{k}_2+\mathbf{k}_3
\mathrm =0$)
is denoted by $\zeta$.  The limits of integration are as follows
\ba
&&s_{123}^-=9 M_{\pi}^2,\qquad s_{123}^+=
(M_{\sigma_{\mathrm III}}- M_{\pi})^2,\nonumber\\
&&s_{12}^-=4 M_{\pi}^2,\qquad s_{12}^=
(\sqrt{s_{123}}-M_{\pi})^2,\nonumber\\
&&s_{34}^\pm=2 M_{\pi}^2+\frac{1}{2 s_{123}}
\left[(s_{123}+M_{\pi}^2-s_{12})(M_{\sigma_{\mathrm III}}^2-M_{\pi}^2-s_{123})\pm\right.\nonumber\\
&&\qquad\left.
\pm\sqrt{\lambda(s_{123},s_{12},M_{\pi}^2)
\lambda(M_{\sigma_{\mathrm III}}^2,s_{123},M_\pi^2)}\right],\nonumber\\
&&s_{23}^\pm=2 M_{\pi}^2+\frac{1}{2 s_{12}}
\left[s_{12}(s_{123}-M_{\pi}^2-s_{12})\right.\pm\nonumber\\
&&\qquad\pm\left.\sqrt{\lambda(s_{12},M_{\pi}^2,M_{\pi}^2)
\lambda(s_{123},s_{12},M_\pi^2)}\right].
\ea
Formula (\ref{4pi}) is similar to that given in \cite{kumar}, however, we
used here different kinematic variables.
As a result, we obtain for the decay into $4\pi$:
\beq
\Gamma_{\sigma_{\mathrm III}\to 2\pi^+2\pi^-}\approx 2.2 \mbox{MeV},\quad
\Gamma_{\sigma_{\mathrm III}\to 2\pi^0\pi^+\pi^-}\approx 1.2 \mbox{MeV},
\quad\Gamma_{\sigma_{\mathrm III}\to 4\pi^0} \approx 0.1 \mbox{MeV}.
\eeq
The total width is
\beq
\Gamma_{\sigma_{\mathrm III}\to4\pi}^{\rm tot}\approx 3.5 \mbox{MeV}
\eeq
and, in the other case ($\sigma_{\mathrm III}\equiv f_0(1710)$),
\beq
\Gamma_{\sigma_{\mathrm III}\to2\pi^+2\pi^-}\approx 6 \mbox{MeV},
\quad\Gamma_{\sigma_{\mathrm III}\to 2\pi^0\pi^+\pi^-}\approx 3.3 \mbox{MeV}
\quad\Gamma_{\sigma_{\mathrm III}\to 4\pi^0}\approx 0.3 \mbox{MeV}.
\eeq
The total width is
\beq
\Gamma_{\sigma_{\mathrm III}\to 4\pi}^{\rm tot}\approx 10 \mbox{MeV}.
\eeq
As one can see, these values are very small.

The other possibility of the state $\sigma_{\mathrm III}$ to decay into
4 pions is to produce two intermediate $\rho$-resonances
($\sigma_{\mathrm III}\to 2\rho\to 4\pi$). Contrary to the decay
through scalar resonances, where strong compensations take place,
in the process with $\rho$-resonances, no compensation
occurs, and it turns out that the decay through $\rho$
determines the most part of the decay width of $\sigma_{\mathrm III}$.

To calculate the amplitude describing the process
$\sigma_{\mathrm III}\to 2\rho$, we need a piece of the Lagrangian
with $\rho$-meson fields. Although we did not consider
vector mesons in the source Lagrangian, an extended
version of NJL model \cite{Volk_86} contains
the vector and axial-vector fields.
Taking  the mass term for $\rho$ mesons from  \cite{Volk_86}
and including dilaton fields into it according to the principle
of scale invariance, we obtain:
\beq
\frac{M_\rho^2}{2}\left(\frac{\chi}{\chi_c}\right)^2(2\rho_\mu^+\rho_\mu^-+
\rho_\mu^0\rho_\mu^0),
\eeq
where $M_\rho=770$
MeV is the $\rho$-meson mass.
From this, we derive the vertex describing
the decay $\sigma_{\mathrm III}\to\rho\rho$:
\beq
\frac{M_\rho^2}{\chi_c}b_{\chi'\sigma_{\mathrm III}}
\chi'(2\rho_\mu^+\rho_\mu^-+
\rho_\mu^0\rho_\mu^0).
\eeq
The decay of a $\rho$-meson into
pions is described by the following amplitude:
\beq
g_\rho (p_1-p_2)^\mu.
\eeq
where $g_{\rho}=6.14$ is the $\rho$-meson decay constant,
$p_1$ and $p_2$ are the momenta of $\pi^+$ and $\pi^-$.
Finally, we come to the following formula for the
amplitude of the process $ \sigma_{\mathrm III}\to\rho^0\rho^0\to 2\pi^+2\pi^-$:
\ba
A_{\sigma_{\mathrm III}\to\rho^0\rho^0\to 2\pi^+2\pi^-}&=&
\frac{M_\rho^2g_{\rho}^2b_{\chi'\sigma_{\mathrm III}}}{\chi_c}
\Biggl((s_{13}+s_{24}-s_{14}-s_{23})
\Delta_\rho(s_{12})\Delta_\rho(s_{34})+\nonumber\\
&+&(s_{13}+s_{24}-s_{12}-s_{34})
\Delta_\rho(s_{14})\Delta_\rho(s_{23})\Biggr),
\ea
 The function $\Delta_\rho(s)$ is the following:
\beq
\Delta_\rho(s)=(s-M_\rho^2+i M_\rho \Gamma_\rho)^{-1}.
\eeq
Here $\Gamma_\rho=150$ MeV is the decay width of the $\rho$ resonance.
The decay into $2\pi^0\pi^+\pi^-$ occurs through a pair of
charged $\rho$-resonances: $\rho^+$ and $\rho^-$. The amplitude of
this process is the same as for the decay with intermediate $\rho^0$.
The decay into $4\pi^0$ cannot go via $\rho$-resonances.

In an extended NJL model \cite{Volk_86}, there are no
vertices describing the decay of a quarkonium  into
$\rho$ mesons. As a result, only the glueball part
determines the decay of $\sigma_{\rm III}$ into 4 pions
through $\rho$ resonances unlike the case with $\sigma$
resonances. This leads to a large decay rate
through $\rho$ mesons (contrary to decays through $\sigma$).

Now let us give the numerical estimates for the decay into 4 pions.
In the case, where $\sigma_{\mathrm III}\equiv f_0(1500)$
 we have:
\beq
\Gamma_{\sigma_{\mathrm III}\to \rho\rho\to 2\pi^+2\pi^-}\approx 50,\quad
\Gamma_{\sigma_{\mathrm III}\to \rho\rho\to2\pi^0\pi^+\pi^-}\approx
90\; \mbox{MeV},
\eeq
with the total width:
\beq
\Gamma_{\sigma_{\mathrm III}\to4\pi}^{\rm tot}\approx 140 \mbox{MeV}.
\eeq
In the other case ($\sigma_{\mathrm III}\equiv f_0(1710)$),
\beq
\Gamma_{\sigma_{\mathrm III}\to\rho\rho\to 2\pi^+2\pi^-}
\approx 350 \mbox{MeV},\qquad
\Gamma_{\sigma_{\mathrm III}\to\rho\rho\to2\pi^0\pi^+\pi^-}
\approx 650\; \mbox{MeV},
\eeq
\beq
\Gamma_{\sigma_{\mathrm III}\to 4\pi}^{\rm tot}\approx 1 \mbox{GeV}.
\eeq

Now we can estimate the total width of the state $\sigma_{\mathrm III}$.
If $\sigma_{\mathrm III}$ is identified with $f_0(1500)$
we have
\beq
\Gamma_{\sigma_{\mathrm III}}^{\rm tot} \approx
220\; \mbox{MeV},
\eeq
which is in qualitative agreement with the experimental
value 112 MeV \cite{PDG},
and, in the other case ($\sigma_{\mathrm III}\equiv f_0(1710)$),
\beq
\Gamma_{\sigma_{\mathrm III}}^{\rm tot}
\approx 1.2\; \mbox{GeV},
\eeq
which is in great contradiction with experimental data.
In the last case ($f_0(1710)$) $\rho$ mesons can show up
as on-mass-shell decay products at large probability.
The decay width is estimated as $\sim$1 GeV. The absence of
this decay mode in experimental observations is a reason
that $f_0(1710)$ is not a glueball.

Our estimates for the decay widths of the scalar meson states
$\sigma_{\mathrm I}$, $\sigma_{\mathrm II}$, and $\sigma_{\mathrm III}$
are collected in Table 3.

\section{Conclusion}
In the  approach presented here, we assume that
(with the exception of the dilaton potential and the 't Hooft
interaction) scale invariance
holds for the effective Lagrangian  before and after SBCS in the chiral limit.
On the other hand, we take into account effects of scale
invariance breaking that come from three sources: the terms
with current quark masses, the dilaton potential reproducing
the scale anomaly of QCD, and term $L_{\rm an}$
induced by gluon anomalies (see (\ref{anomalynotscaled})
in the Introduction).

The scale invariance breaking that is connected with the
term  $L_{\rm an}$ was not taken
into account in our previous paper \cite{EPJA2000}%
\footnote{Note that there was wrong sign at the term
in formula (43) that describes the quarkonia-glueball mixing,
which led to incorrect estimates for the decay widths of
the scalar glueball.}.
This led to a small quarkonia-glueball mixing proportional to current
quark masses, disappearing in the chiral limit.
If the term $\Delta L_{\mathrm an}$ taken into account
in (\ref{Lagr:bosonized:chi})
the quarkonia-glueball mixing becomes much greater and
does not disappear in the
chiral limit, being proportional to constituent quark masses
(quark condensates). This corresponds to results obtained
from QCD in \cite{shifman}. This contribution to
the quarkonia-glueball mixing turns out to have decisive
effect on the strong decay widths of
scalar mesons.

For the scalar meson states $f_0(400-1200)$ and $f_0(980)$,
we obtain good agreement with experimental data \cite{PDG}.
Their decay widths are determined by quarkonium parts of
decay amplitudes.

Strong decays of the scalar meson state $\sigma_{\mathrm III}$
("glueball") are considered for two different masses:
1500 MeV and 1710 MeV.
In the $\pi\pi$ channel, the contribution from quarkonia prevails
over that from the glueball and thereby determines the
decay rate.
In the case of $KK$, $\eta\eta$, $\eta\eta'$, and $\pi\pi$
channels, there are noticeable compensations
among decay amplitudes.

A similar situation with compensations takes place in
the decay into 4$\pi$ with intermediate
$\sigma$-mesons. Here we have a strong compensation
among the glueball and quarkonia contributions. But
there is a possibility for the state $\sigma_{\mathrm III}$
to decay through $\rho$-resonances.
In this case, as the quarkonium component is absent,
no compensation occurs, and this channel
determines the most of the total decay width of
$\sigma_{\mathrm III}$.

We performed calculations for both candidates for
the scalar glueball state: $f_0(1500)$ and $f_0(1710)$,
and found that $f_0(1500)$ is rather the glueball.
The main decay mode is that into 4 pions.
The decay rate into a pair of kaons is next by order
of magnitude and is followed by the  $\eta\eta$,
$\eta\eta'$, and $\pi\pi$ decay modes.

The total width of the third scalar isoscalar state is estimated to
be about 220 MeV for $M_{\sigma_{\mathrm III}}=1500$ MeV and
1.2 GeV for $M_{\sigma_{\mathrm III}}=1710$ MeV.
The experimental width of $f_0(1500)$  is 112 MeV
and that of $f_0(1710)$ is 130 MeV.
 Unfortunately, the
detailed data on the branching ratios of $f_0(1500)$
and $f_0(1710)$ are not reliable and are
controversial \cite{PDG}.

Our calculations are rather qualitative. However,
they allow us to conclude that $f_0(1500)$ is
a scalar glueball state, whereas
$f_0(1710)$ is a quarkonium, for the following reasons:
1) The total decay width of the glueball in our model
better fits to its experimental value if
 $f_0(1500)$ is assumed to be the glueball, rather than
 $f_0(1710)$. 2)As it follows from our calculations,
 the main decay mode of the scalar
 glueball is that into four pions.
 This is true for the state $f_0(1500)$. A decay
of $f_0(1710)$ into four pions, however, was not
seen in experiment. 3) Moreover, a direct decay of
into a pair of $\rho$ mesons on
their mass-shell is possible for a scalar glueball
with the mass about 1.7 GeV. It has also not been
observed.
Our conclusion concerning  the nature of $f_0(1710)$
as a quarkonium state is in agreement with the
conclusion made in our papers \cite{ECHAJA1999}.

We are going to use this approach in our future work
for  describing both glueballs and ground and radially excited scalar
meson nonets which lie it the energy interval from 0.4 to 1.71 GeV.

\section*{Acknowledgement}
We are grateful to Profs.~A.A.~Andrianov, D.~Ebert and Drs.
~A.E.~Dorokhov, S.B.~Gerasimov, and  N.I.~Kochelev
for useful discussions. The work is
supported by RFBR Grant 00-02-17190.
\bigskip

\clearpage

\centerline{{\bf\sc references}}

\clearpage
\section{Table captions}
\begin{enumerate}
\item[\textbf{Table \ref{T:spectr}.}] The masses of physical scalar meson states $\sigma_{\mathrm I}$,
$\sigma_{\mathrm II}$, $\sigma_{\mathrm III}$ and the values of  the parameters $\chi_c$,
$\chi_0$, bag constant $B$, and (bare) glueball mass $m_g$ (in MeV)
for two cases: 1) $M_{\sigma_{\mathrm III}}=1500$ MeV and
2) $M_{\sigma_{\mathrm III}}=1710$ MeV
\item[\textbf{Table \ref{mix}.}] Elements of the matrix $b$,
describing mixing in the scalar isoscalar sector.
The upper table refers to the case  $\sigma_{\mathrm III}\equiv f_0(1500)$,
the lower one to the case  $\sigma_{\mathrm III}\equiv f_0(1710)$
\item[\textbf{Table \ref{Gdecays}.}]
The partial and total decay widths (in MeV)
of the scalar meson states $f_0(400-1200)$,
$f_0(980)$ and of the glueball for two cases:
$\sigma_{\mathrm III}\equiv f_0(1500)$ and $\sigma_{\mathrm III}\equiv f_0(1710)$, and experimental
values of decay widths of $f_0(1500)$ and $f_0(1710)$ \protect\cite{PDG}.

\end{enumerate}

\clearpage
\section{Tables}

\begin{table}[p]
\centering
\caption{}
\label{T:spectr}
\begin{tabular}{||c|c|c|c|c|c|c|c||}
    \hline
    & $\sigma_{\mathrm I}$ & $\sigma_{\mathrm II}$ & $\sigma_{\mathrm III}$ & $\chi_c$ & $\chi_0$ &$B, \mbox{GeV}^4$& $M_g$\\ \hline
  I & 400 & 1100 & 1500 & 206 & 190 & 0.009 & 1447\\
  II & 400 & 1100 & 1710 & 180 & 166 & 0.009 & 1665\\ \hline
\end{tabular}
\end{table}

\begin{table}[p]
\caption{}
\label{mix}
\centering

\begin{tabular}{|c|rrr|}
\hline
 & $\sigma_{\mathrm I}$ & $\sigma_{\mathrm II}$ & $\sigma_{\mathrm III}$\\
\hline
$\sigma_u^r$ & 0.939 & 0.240 & $0.247$ \\
$\sigma_s^r$ & $-0.214$ & 0.968 & $-0.128$ \\
$\chi'$ & $-0.270$ & 0.067 &  0.960 \\
\hline
\end{tabular}
\begin{tabular}{|c|rrr|}
\hline
 & $\sigma_{\mathrm I}$ & $\sigma_{\mathrm II}$ & $\sigma_{\mathrm III}$\\
\hline
$\sigma_u^r$ & 0.948 & 0.232 & $0.216$ \\
$\sigma_s^r$ & $-0.216$ & 0.971 & $-0.099$ \\
$\chi'$ & $-0.233$ & 0.047 &  0.971 \\
\hline
\end{tabular}
\end{table}

\begin{table}[p]
\caption{}
\label{Gdecays}
\centering\begin{tabular}{||r|r|r|r|r|r|r|r||}
\hline
& $\Gamma_{\pi\pi}$ & $\Gamma_{K \bar K}$ & $\Gamma_{\eta\eta}$ & $\Gamma_{\eta\eta'}$ &
$\Gamma_{4\pi}$ & $\Gamma_{\mathrm tot}$ & $\Gamma_{\mathrm tot}^{\mathrm exp}$\\
\hline
$f_0(400-1200)$ & 820 & -- & -- & --  & -- & 820 & 600--1200\\
$f_0(980)$ & 28 & -- & -- & -- & --  & 28 & 40--100\\
$f_0(1500)$ & 14 & 28 & 25 & 13 & $\sim$ 140 & $\sim$ 220 & 112 \\
$f_0(1710)$ & 8 & 60 & 43 & 31 & $\sim$ 1000 & $\sim$ 1100 & 130 \\
\hline
\end{tabular}
\end{table}

\end{document}